\documentclass[aps,prl,twocolumn,superscriptaddress]{revtex4-2}
\usepackage{amsfonts}
\usepackage{amsmath}
\usepackage{mathrsfs}
\usepackage{amssymb}
\usepackage{graphicx}
\usepackage{lipsum} 
\usepackage[T1]{fontenc}
\usepackage{epstopdf}
\usepackage{bm}
\usepackage{color}

\usepackage[colorlinks=true,linkcolor=blue,anchorcolor=blue,citecolor=blue,urlcolor=blue]{hyperref}

\newcommand{\rev}[1]{\textcolor{black}{{#1}}}

\begin{document}

\title{Topological pumping of multi-frequency solitons}

\author{Yaroslav V. Kartashov }
\email{kartashov@isan.troitsk.ru}
\affiliation{
	Institute of Spectroscopy, Russian Academy of Sciences, Troitsk, Moscow, 108840, Russia}

\author{Fangwei Ye} 
\email{fangweiye@sjtu.edu.cn}
\affiliation{School of Physics and Astronomy, Shanghai Jiao Tong University, Shanghai 200240, China}

\author{Vladimir V. Konotop}
\email{vvkonotop@ciencias.ulisboa.pt}
\affiliation{
	Departamento de F\'isica and Centro de F\'isica Te\'orica e Computacional, Faculdade de Ci\^encias, Universidade de Lisboa, Campo Grande, Edif\'icio C8, Lisboa 1749-016, Portugal 
}

\begin{abstract}
We report on the topological pumping of quadratic optical solitons, observed through their quantized transport in a dynamic optical potential. A distinctive feature of this system is that the two fields with different frequencies, which together form the quadratic soliton, evolve in separate yet topologically equivalent dynamic optical potentials. Pumping in this system exhibits several notable differences from pumping in cubic media. While Chern indices characterizing quantized transport for uncoupled fundamental and second harmonic waves are nonzero, small-amplitude solitons with narrow spectra do not move, thus revealing a non-topological phase. \rev{As the nonlinearity increases,} the system undergoes a sharp transition, depending on the velocity of one of \rev{the} sublattices forming dynamical potential, into the phase where the quantized transport of quadratic solitons governed by nonzero Chern numbers is observed. The power level at which this transition occurs increases with increase of pumping velocity, \rev{and the} transition is observed even in the regime \rev{when} the adiabatic approximation \rev{no longer applies}. Unlike in cubic media, in \rev{a} quadratic medium neither breakup of topological pumping nor fractional pumping at high power levels are observed.

\end{abstract}

\maketitle

\rev{ Quantized transport refers to the transfer of a physical quantity driven by periodic pumping, where the amount transferred in each cycle takes on discrete values. These values are determined by global properties of the system and are insensitive to its local parameters.} Discovered by Thouless~\cite{Thouless} for electrons in a periodic potential, \rev{it is now known} to be  a ubiquitous phenomenon encountered in diverse physical systems~\cite{Citro2023}. It is commonly observed in wave processes occurring in media, whose parameters vary adiabatically and periodically. Being linear in nature, such pumping is topological, meaning that it can be characterized by the topological indices defined in the coordinate-time space. The possibilities of experimental realization of pumping in optical~\cite{Zilberberg2018, Wang2022, Jurgesen2021} and atomic~\cite{Lohse2016, Nakajima2016, Taddia2017, Lohse2018, Nakajima2021} systems, where dynamically varying periodic potentials can be routinely created and where the nonlinear response of the medium is a naturally present factor, raises the important question about the impact of nonlinearity on the pumping process. This problem is challenging for several reasons. The definition of topological (specifically, Chern) indices usually cannot be unambiguously extended to nonlinear media and require constraints on specific solutions used for calculation of these indices~\cite{Maczewsky2020, Jurgesen2021, Sone2024} that may not be universal. Nonlinearity is a source of various instabilities, in particular of linear~\cite{KonSal} and  nonlinear~\cite{Bronski2001} Bloch waves, significantly complicating their experimental realization and necessitating the approaches extending beyond the concept of nonlinear Bloch waves. Nonlinearity, when it balances dispersion or diffraction, can lead to the formation of self-sustained nonlinear excitations - lattice solitons - whose parameters depend on their amplitude. Ultimately, the nonlinearity results in the interchange of energy (or particles) among different bands of periodic potential~\cite{Fu2022,Fu2022a}, leading to varying band populations upon topological pumping and invalidating the approximation of completely filled bands.

The above mentioned nonlinear effects may have dramatic impact on the topological pumping and introduce unusual features into this process. While first experiments illustrated the remarkable fact that topological pumping can be realized with solitons at low and intermediate power levels in cubic medium \cite{Jurgesen2021}, the possibility of breakup of quantized transport at high powers and fractional pumping were also revealed~\cite{Nakagawa2018, Jurgesen2021, Fu2022, Fu2022a, Jurgesen2023}, as well as the emergence of nonlinearity-induced topological phases in systems, where transport does not exist in linear and even weakly-nonlinear regimes~\cite{Maczewsky2020, Kuno2020}. This hints on the fact that the strength of nonlinear processes and the very type of nonlinear response of the medium can significantly and qualitatively affect the properties of quantized topological transport.

Parametric optical processes involving multi-frequency fields are important example of such nonlinear processes from both theoretical and applied perspectives, which have not been investigated, so far, from the perspective of quantized transport. Unlike in previously considered cubic nonlinear media, where topological transport is altered by self-action of single-frequency (single-component) excitations, parametric systems are characterized by nonlinear coupling of several fields (two, in the case of this Letter), which "feel" effectively different dynamical potentials. Thus, the transport properties of the components are \textit{a priori} different locally, but may become globally unified by the parametric interactions. In this Letter \rev{through numerical study of a model describing quadratic solitons,} we show that topological transport of solitons is possible in optical materials with quadratic nonlinearity, where it acquires unusual features, sharply different from transport in cubic materials. \rev{We also describe transition between topological and non-topological transport, controlled by nonlinearity.}

We consider transport of one-dimensional solitons in a $\chi^{(2)}$ optical medium governed by the equations \cite{Torner2002, Buryak2002} 
\begin{align}
 	\label{main1}
 	i\frac{\partial \Psi_1}{\partial z}=-\frac{1}{2}\frac{\partial^2 \Psi_1}{\partial x^2}-V(x,z)\Psi_1-\Psi_1^*\Psi_2  
 	\\
 	\label{main2}
 	i\frac{\partial \Psi_2}{\partial z}=-\frac{1}{4}\frac{\partial^2 \Psi_2}{\partial x^2} +\beta \Psi_2-2V(x,z)\Psi_2-\Psi_1^2 
\end{align}
Here $\Psi_1$ and $\Psi_2$ are the normalized \rev{complex-valued} amplitudes of the fundamental frequency (FF) and second harmonic (SH) fields, respectively, $\beta$ is the \rev{real} phase mismatch, $V(x,z)=V_1(x)+V_2(x-\alpha z)$ is the optical potential created by two periodic sublattices $V_j(x)=p_j\cos^2(\pi x/d_j)$ with \rev{real} depths $p_{1,2}$ and commensurate periods $d_{1,2}$, and  $\alpha\ll 1$ is the sliding angle (or velocity) of the second sublattice with respect to the first one. Respectively $X=\max\{d_1,d_2\}$ and $Z=X/\alpha$ are the transverse and longitudinal periods of the potential. We adopt normalizations as in \cite{Kartashov2004, Kartashov2021}.

Quasi-stationary nonlinear modes that adiabatically follow the change of the optical potential can be searched in the form $ \Psi_j=\psi_j(x,z)\exp[ij\int_0^zb(\zeta)d\zeta]$, where $j=1,2$, $b(z)$ is a smoothly varying function, and $\psi_j(x,z)$ solve
\begin{align}
	\label{main-stat-1}
	b\psi_1=-H_1\psi_1+\psi_1^*\psi_2,  
	\quad
	b\psi_2=-H_2\psi_2+\frac{1}{2}\psi_1^2.  
\end{align}
Here $H_{1,2}$ can be obtained by deformation (via variation of $\epsilon$) of the Hamiltonian $H_{\epsilon}=-{1}/{(2\epsilon^2)}\partial_x^2 + {(\epsilon-1)}\beta/{2}-V(x,z)$ at $\epsilon= 1,2$. \rev{In the adiabatic regime}, the terms with $db(z)/dz$ and $d\psi_j/dz$ were dropped and hereafter we omit the explicit $z$-dependence of all adiabatically varying parameters, except for the potential. When potential $V$ is static, i.e., does not depend on $z$, as it happens for example for $\alpha=0$, Eq.~(\ref{main-stat-1}) gives rise to rich families of quadratic lattice solitons, studied in both continuous and discrete systems in \cite{Bang1997, Peschel1998, Kobyakov1999, Sukhorukov2000, Malomed2002, Iwanow2004, Xu2005, Susanto2007, Setzpfandt2009, Setzpfandt2010}.

\begin{figure}
    \includegraphics[width=\columnwidth]{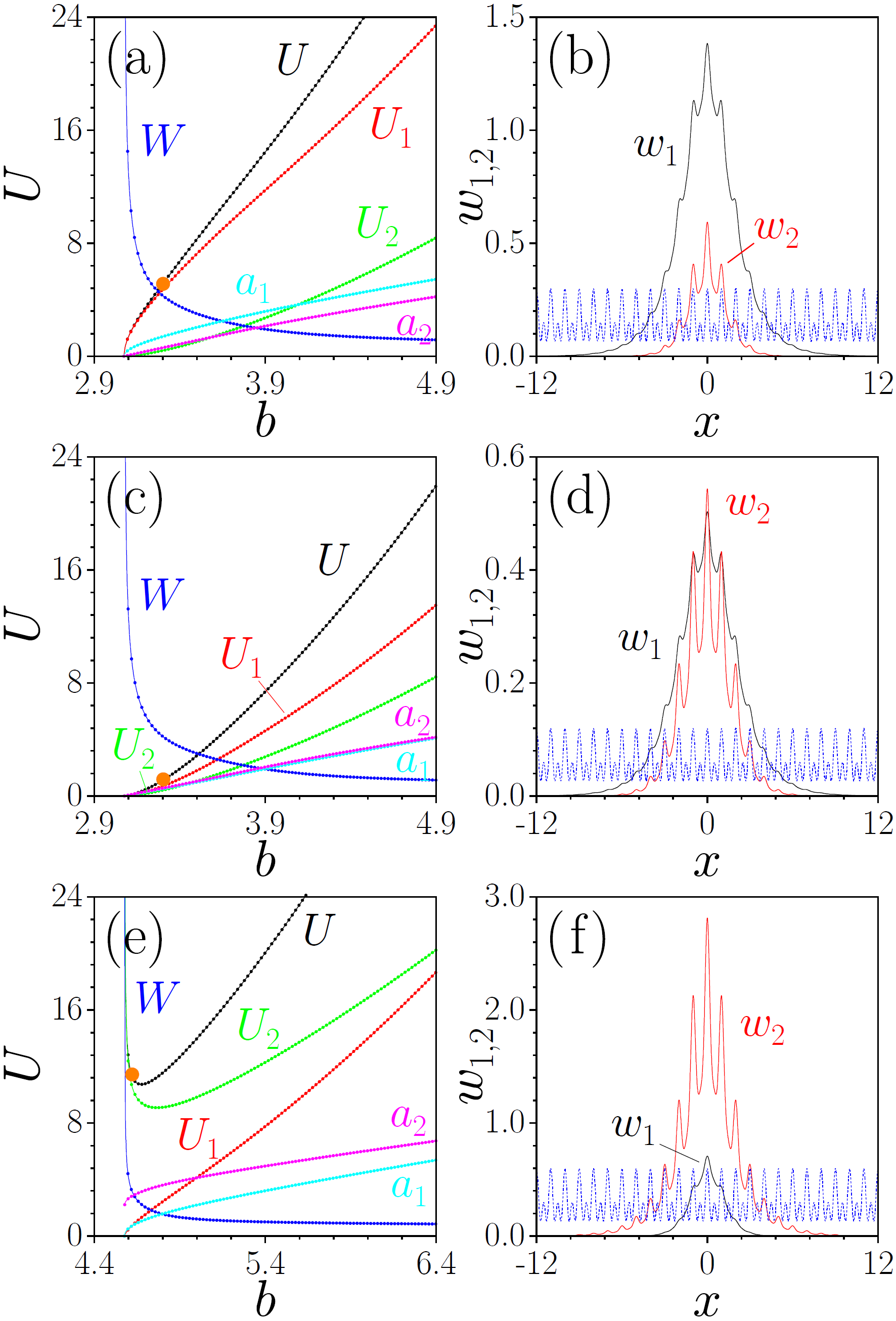}
    \caption{Total soliton power $U$, integral soliton width $W$, amplitudes $a_{1,2}$ and powers $U_{1,2}$ of FF and SH components versus propagation constant $b$ at $\beta=+3.5$ (a), $\beta=+0.5$ (c), and $\beta=-2.5$ (e). Panels (b), (d), (f) show soliton profiles corresponding to the orange dots in (a), (c), (e), respectively. Here and in all figures below $p_{1,2} =3$, $d_1 =0.5$, $d_2 =1$. For these parameters $\beta_{\rm cr}\approx 0.5$.}
    \label{fig:one}
\end{figure}

Since solitons are localized and $\psi_{1,2}$ vanish as $|x|\to\infty$, the linear properties of the Hamiltonians $H_{1}$ and $H_2$ become particularly important. Specifically, the \rev{difference between the locations} of the upper gap edges of $H_{1}$ and $H_2$ determines the properties of quadratic solitons supported by these lattices~\cite{Moreira2012, Moreira2013}. Let $\mathfrak{b}_j(k)$ be the Bloch spectrum of the Hamiltonian $H_j$ at $z=0$ with the Bloch wavenumber $k$ in the reduced Brillouin zone (BZ): $k\in [-{\rm K}/2,{\rm K}/2)$ where  $\textrm{K}=2\pi/X$. Thus, $H_j\varphi_j=\mathfrak{b}_j(k)\varphi_j$, where $\varphi_j=e^{ikx}u_{jk}(x)$ and $u_{jk}(x)=u_{jk}(x+X)$ (we omit the band index since only the upper band is considered here). Focusing on solitons in the semi-infinite gap, we denote by $b_{1,2}=\mathfrak{b}_{1,2}(k=0)$ the lower boundaries of the semi-infinite gaps in spectra of $H_{1,2}$. In the chosen parametrization $b_1$ is $\beta$-independent, while $b_2$ linearly decreases with $\beta$. \rev{Let us define a critical value of the phase mismatch $\beta_{\rm cr}$ at which $b_1(0)=b_2(0)$ is achieved. Then} one can distinguish three cases~\cite{Moreira2012,Moreira2013}: (i) $b_2>b_1$ (that is achieved at $\beta<\beta_{\rm cr}$), (ii) $b_2=b_1$ ($\beta=\beta_{\rm cr}$), and (iii) $b_2<b_1$ ($\beta>\beta_{\rm cr}$). Quadratic solitons \rev{can} exist \rev{only} at $b>b_{\rm co}=\max\{b_1,b_2\}$ \rev{(the subindex "co" standing for cutoff; see the illustration in  Fig.~S1 of Supplemental Material~\cite{SupplMat}).} Soliton families (supported by the lattice taken at $z=0$) for these three cases are shown in Fig.~\ref{fig:one} in the first, second and third rows, respectively. The essential differences are observed in the behavior of soliton families near the gap edge in panels (a) and (e). While in (a) at $\beta>\beta_{\rm cr}$ \rev{(in this case $b_{\rm co}=b_1$)} the total soliton power $U(b)=U_1+U_2$, where $U_j=\int |\Psi_j|^2dx$, vanishes at $b\to b_\textrm{co}$, in panel (e) at $\beta<\beta_{\rm cr}$ \rev{(where $b_{\rm co}=b_2$)} one has $U(b)>U_{\rm th}>0$ and therefore solitons exist only above threshold power $U_{\rm th}$. This is reflected also in relations between amplitudes of FF and SH waves in these three regimes: (i) $a_1>a_2$, (ii) $a_1\approx a_2$, and (iii) $a_1<a_2$. These last properties are obvious when one compares soliton profiles shown in panels (b), (d), and (f) that correspond to the orange dots in $U(b)$ curves. In all cases the integral soliton width $W=[(1/U)\int x^2(|\Psi_1|^2+|\Psi_2|^2)dx]^{1/2}$ diverges as $b\to b_\textrm{co}$.  

\begin{figure}
    \includegraphics[width=\columnwidth]{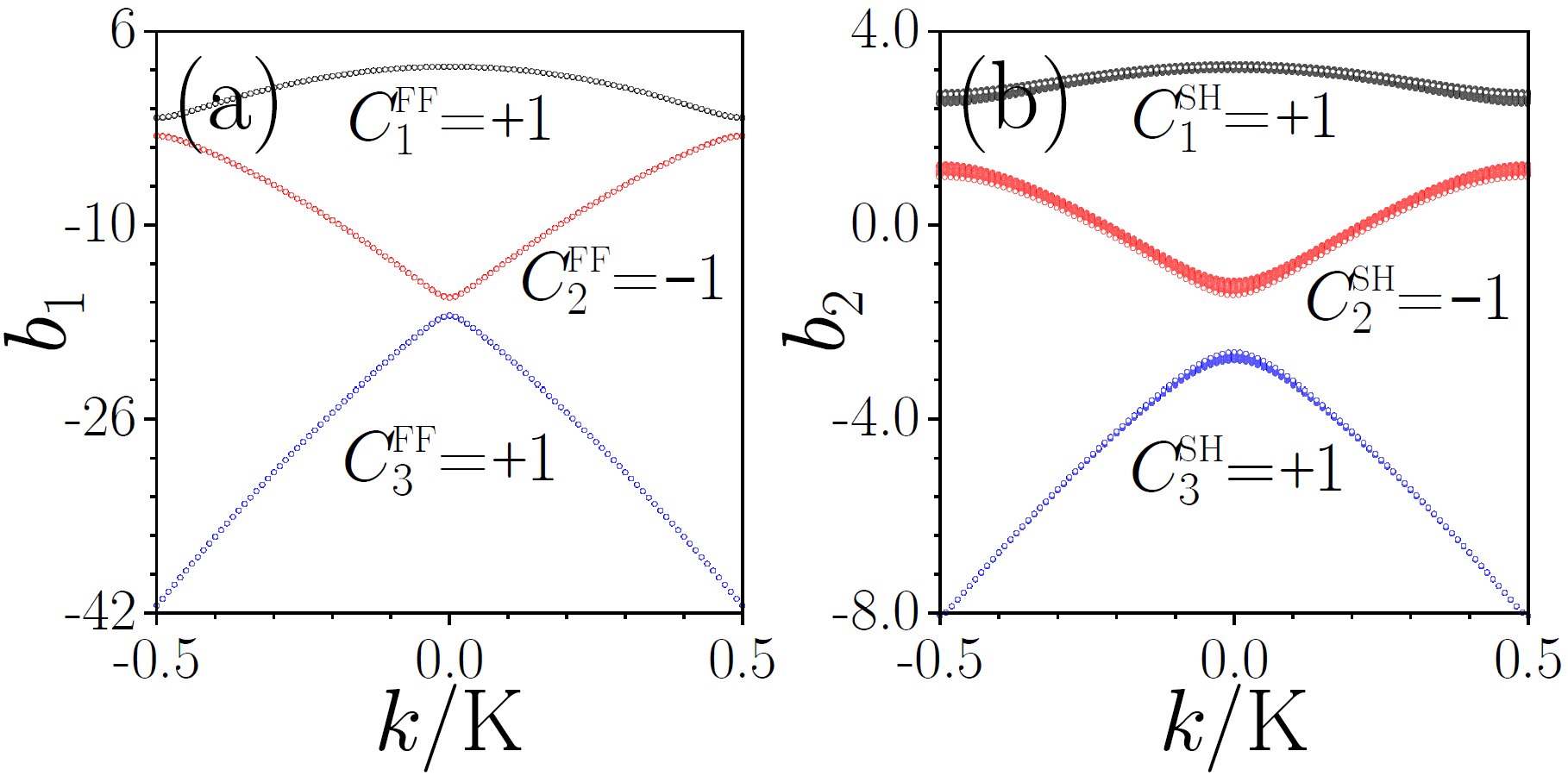}
    \caption{One-cycle evolution of the first three bands for FF (a) and SH (b) waves for $\beta=0$ (bands at different distances $z$ are superimposed). Corresponding Chern numbers are indicated near each band. 
}
    \label{fig:two}
\end{figure}

The link between the Hamiltonians $H_1$ and $H_2$ that can be obtained via deformation (homotopy) of $H_\epsilon$ implies that the topological properties of $H_{1,2}$ are identical, even though their local properties are different. Specifically, their bands are encountered in a one-to-one correspondence, and the space-"time" Chern numbers $C_\nu=(i/2\pi)\int_0^Zdz\int_{\rm BZ}dk[\langle \partial_zu_{jk}|\partial_ku_{jk}\rangle -\langle \partial_ku_{jk}|\partial_zu_{jk}\rangle]$ (\rev{here $\langle f|g\rangle \equiv\int f^*gdx$ and} the role of "time" is played by the coordinate $z$) of the $\nu$-th bands are equal for both Hamiltonians, as illustrated in Fig.~\ref{fig:two}. Notice that "instantaneous" bands shown in Fig.~\ref{fig:two} for all distances within one period $Z$ change with $z$ because potential $V(x,z)$ varies, but bands with different $\nu$ never overlap. \rev{(The Chern numbers $\pm 1$ in  Fig.~\ref{fig:two} are specific to the chosen potential but can be adjusted to other integers by modifying the lattice.)} Thus, in the absence of nonlinearity and in accordance with Thouless prediction, the quantized pumping of fields $\psi_{1,2}$ populating bands with the same $\nu$ should result in equal one-cycle shifts of 
the center of mass of the beam, $x_c(z)=(1/U)\int x(|\Psi_1|^2+|\Psi_2|^2)dx$.   

A peculiarity of the pumping of quadratic solitons of small amplitude is that $b_2(z)-b_1(z)$ can change its sign over one longitudinal period $Z$. This certainly happens at the critical phase mismatch $\beta=\beta_{\rm cr}$ when $b_2(0)=b_1(0)$. This means that during one-cycle evolution the small-amplitude quadratic solitons [similar to the one shown in Fig.~\ref{fig:one} (d)] cannot be driven adiabatically. In other words, a continuous transition to linear Thouless pumping in such a system is impossible (quantized transport occurs only when the nonlinearity is equal to zero). Therefore, in what follows, we focus on the cases where $\beta$ is sufficiently large for the sign of $b_2(z)-b_1(z)$ to remain constant during the entire evolution. 

\begin{figure}
    \includegraphics[width=\columnwidth]{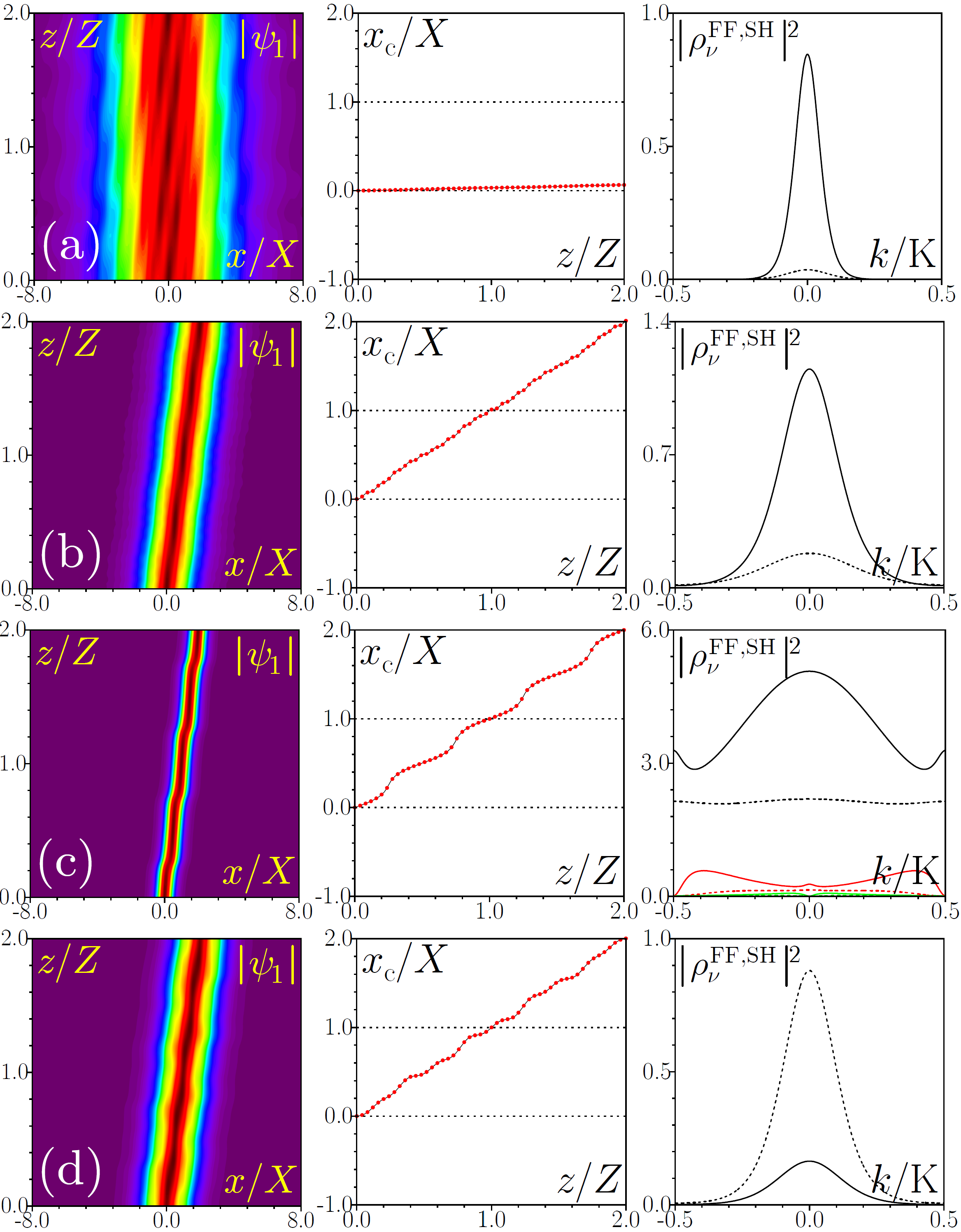}
    \caption{Pumping of quadratic solitons. Each row shows (from left to right) propagation dynamics of the FF component $\psi_{1}$ (dynamics of SH component is similar), the evolution of the integral soliton center $x_c(z)$ over two pumping cycles, and projections $|\rho_{\nu}^{\rm FF,SH}|^2$ of soliton profile on Bloch bands at $z=0$. The solid and dashed lines correspond to projections for FF and SH waves, respectively. The black lines show projections on $\nu=1$ band, red lines that become visible for narrow soliton in (c) show projections on $\nu=2$ band, the green line in (c) is the projection on $n=3$ band. Input solitons correspond to (a) $b=3.25$, $\beta=+3.5$, (b) $b=4$, $\beta=+3.5$ (c) $b=12$, $\beta=+3.5$, (d) $b=4.75$, $\beta=-2.5$. In all cases $\alpha=0.01$, $X=1$.}
    \label{fig:three}
\end{figure}

In Fig.~\ref{fig:three} we show several characteristic regimes of the quantized transport of quadratic solitons, when phase mismatch $\beta$ is taken sufficiently far from the critical value $\beta_{\rm cr} \approx 0.5$. The first regime that appears counter-intuitive at the first glance is illustrated in line (a), where a small-amplitude broad soliton (taken close to the cutoff $b_\textrm{co}$) does not undergo any appreciable displacement over one cycle of pumping, although it bifurcates from the first band [Fig.~\ref{fig:one}(a)] that is topologically nontrivial: the Chern numbers for both FF and SH waves are $+1$ (Fig.~\ref{fig:two}). This effect becomes clear when one considers solitonic nature of the initial wave packet. Unlike strongly localized linear wave packets, which have spectra as wide as the BZ (corresponding to nearly fully populated top band), a small-amplitude soliton has sufficiently narrow spectrum. This wavepacket is an envelope of the Bloch mode corresponding to $\mathfrak{b}_j(0)$, for which $\partial \mathfrak{b}_j/\partial k=0$, indicating zero group velocity. Thus, this situation is opposite to the one in which quantized transport is observed: the soliton spectrum [see the right panel in Fig.~\ref{fig:three}(a), where we show projections $|\rho_{\nu}^{\rm FF,SH}|^2$ of the soliton on Bloch bands] does not fill the whole allowed band, which is necessary for quantized transport. 
Thus, the absence of transport at sufficiently small soliton amplitudes in a  $\chi^{(2)}$ 
  medium has a very different physical origin compared to the previously reported examples of transport breakdown at high amplitudes in cubic media~\cite{Jurgesen2021, Fu2022}.

When one moves farther from the cutoff $b_\textrm{co}$, nonlinearity becomes stronger, the width of the soliton decreases and its spectrum broadens. Remarkably, in this case and for $\beta>\beta_\textrm{cr}$ after a narrow transition region (see Fig.~\ref{fig:four} below) one observes quantized transport of quadratic solitons that is exceptionally robust. In this regime the one-cycle shift of the soliton center is given by $x_c(Z)=X$, i.e., it coincides with Thouless prediction~\rev{\cite{Thouless}} for the shift of linear wavepacket, whose spectrum is wider than the spectral width of the upper band (having Chern index $+1$). This behavior of quadratic solitons resembles the quantized transport of solitons in a Kerr medium, indicating that the phenomenon of quantized soliton transport is universal and can occur in a wide range of materials. However, sharp differences in pumping dynamics of quadratic and cubic solitons are observed at even larger soliton powers $U$. While high-power solitons in cubic medium exhibit either breakup of pumping or fractional pumping~\cite{Fu2022}, neither of these phenomena are observed for quadratic solitons that exhibit robust pumping even for very large values of $U$ (large amplitudes), as illustrated in Fig.~\ref{fig:three}(c). This effect can be attributed to the suppression of power exchange between upper and lower bands that is observed for both FF and SH waves [see right panel in Fig.~\ref{fig:three}(c)]. Quantized transport is also observed for solitons in the regime $\beta>\beta_\textrm{cr}$, where solitons exist above power threshold $U_\textrm{th}$, as illustrated in Fig.~\ref{fig:one}(e),(f) \rev{(the results obtained for even larger phase mismatch $\beta\gg 1$, known as the cascading limit are presented in~\cite{SupplMat})}. The dynamics shown in Fig.~\ref{fig:three}(d) displays the quantized transport of quadratic soliton with dominating SH component according to the prediction $x_c(Z)=X$.

To understand the universality of quantized transport of finite-amplitude quadratic solitons, in the left column of Fig.~\ref{fig:one} we observe two characteristic properties of finite-amplitude solitons: for $b$ large enough [larger than $3.7$ in Figs.~\ref{fig:one}(a) and (c), and larger than $4.8$ in Fig.~\ref{fig:one}(c)] the widths of solitons do not change much. \rev{The soliton widths} are estimated as $W\lesssim 1.5$, i.e., they are narrower than two periods of the lattice, while the amplitude of the SH can be well approximated by a linear function of the propagation constant $b$. This suggests that in the leading order one can approximate $\Psi_1\approx e^{ibz}A_1(z)w_1(x,z)$ and $\Psi_2\approx e^{2ibz}A_2(z)w_2(x,z)$, where $A_{1,2}$ are the amplitudes of the fields and $w_{1,2}(x,z)$ are the functions describing local spatial distributions of the modes. In order to clarify the meaning of the spatial distributions, we recall that the dependence of $A_{j}$ and $w_j$ on $z$ is adiabatic, and in the leading approximation $z$ in $H_{1,2}$ is considered as a parameter. Thus, the functions $w_{1,2}(x,z)$ for a given $z$ should be considered as elements of some orthonormal bases. Considering required spatial localization of $w_{1,2}(x,z)$ it is natural to choose them as Wannier functions~\cite{Kohn1959} of $H_1$ and $H_2$, which are centered in the same spatial location because both Hamiltonians include the same optical potential $V(x,z)$. Without loss of generality we can consider the Wannier functions which for $z=0$ are centered at $x=0$. Furthermore, for relatively deep lattices  in the leading order one can approximates $H_jw_j\approx -b_jw_j$ omitting the terms related to the hopping between neighboring lattice sites and to interband transitions (see e.g. ~\cite{AKKS,Fu2022}). Subject to these assumptions, substituting the ansatz $\Psi_j\approx e^{ibz}A_jw_j$ in (\ref{main-stat-1}), projecting the FF and SH equation over $w_1$ and $w_2$, and introducing $a_j=\chi A_j$, where $\chi_j=\int w_1^2w_2^*dx$, we obtain $a_2=b-b_1$, $a_1^2=2(b-b_1)(b-b_2)$. 

The described approximation reproduces qualitative behavior of soliton families in Fig.~\ref{fig:one}. It also allows for understanding of quantized transport, because one-cycle displacement of the center of mass of an optical beam $x_c(z)$ with profile corresponding to the Wannier function of the upper band with $C_1=+1$ is $x_c(Z)=X$~\cite{Fu2022}. In other words, the quantized transport at large amplitudes can be viewed as a Thouless pumping of quadratic Wannier solitons (for Wannier solitons in  cubic media see~\cite{AKKS}).

Returning to the dynamics illustrated in Fig.~\ref{fig:three} at $\beta>\beta_\textrm{cr}$, one can distinguish two different phases of the adiabatic pumping of quadratic solitons: a {\em non-topological phase at low powers}, where soliton shift is nearly zero, and {\em a topological phase at sufficiently high powers}, where shift is quantized, separated by a narrow phase transition region. Since at $\beta>\beta_\textrm{cr}$ the power monotonically increases with propagation constant $b$, this phase transition is clearly visible in the dependence of the one-cycle shift of soliton center on $b$ illustrated in Fig.~\ref{fig:four}(a) by the black line corresponding to slow modulation $\alpha =0.001$. The non-topological phase extends in this case over a finite interval of propagation constants: $b\in(b_{\rm co},b_{\rm tr})$. It should be mentioned that this interval gradually becomes narrower when $\alpha\to 0$, even though it never vanishes at $\beta>\beta_\textrm{cr}$. For instance, below  $\alpha =0.001$ further decrease of sliding velocity does not lead to dramatic decrease of this interval. In contrast, at $\beta<\beta_\textrm{cr}$, where solitons exist only above power threshold $U_\textrm{th}$, this phase transition behaves quite differently near the cut-off value $b_{\rm co}$, since the region of the propagation constants $(b_{\rm co},b_{\rm tr})$, where non-topological phase is observed in this case, shrinks practically to zero as $\alpha\to 0$ [see black curve in Fig.~\ref{fig:four}(b)]. Remarkably,  increase of the pumping velocity enlarges the interval $(b_{\rm co},b_{\rm tr})$, where non-topological phase is observed and makes transition region broader, but it does not affect the very fact of existence of two different phases, see red and blue curves in Fig.~\ref{fig:four}(a) and (b). Notice that at sufficiently high $b$ values pumping of quadratic solitons is observed even at $\alpha=0.05$, in the regime that is far from adiabatic one.

\begin{figure}
    \includegraphics[width=\columnwidth]{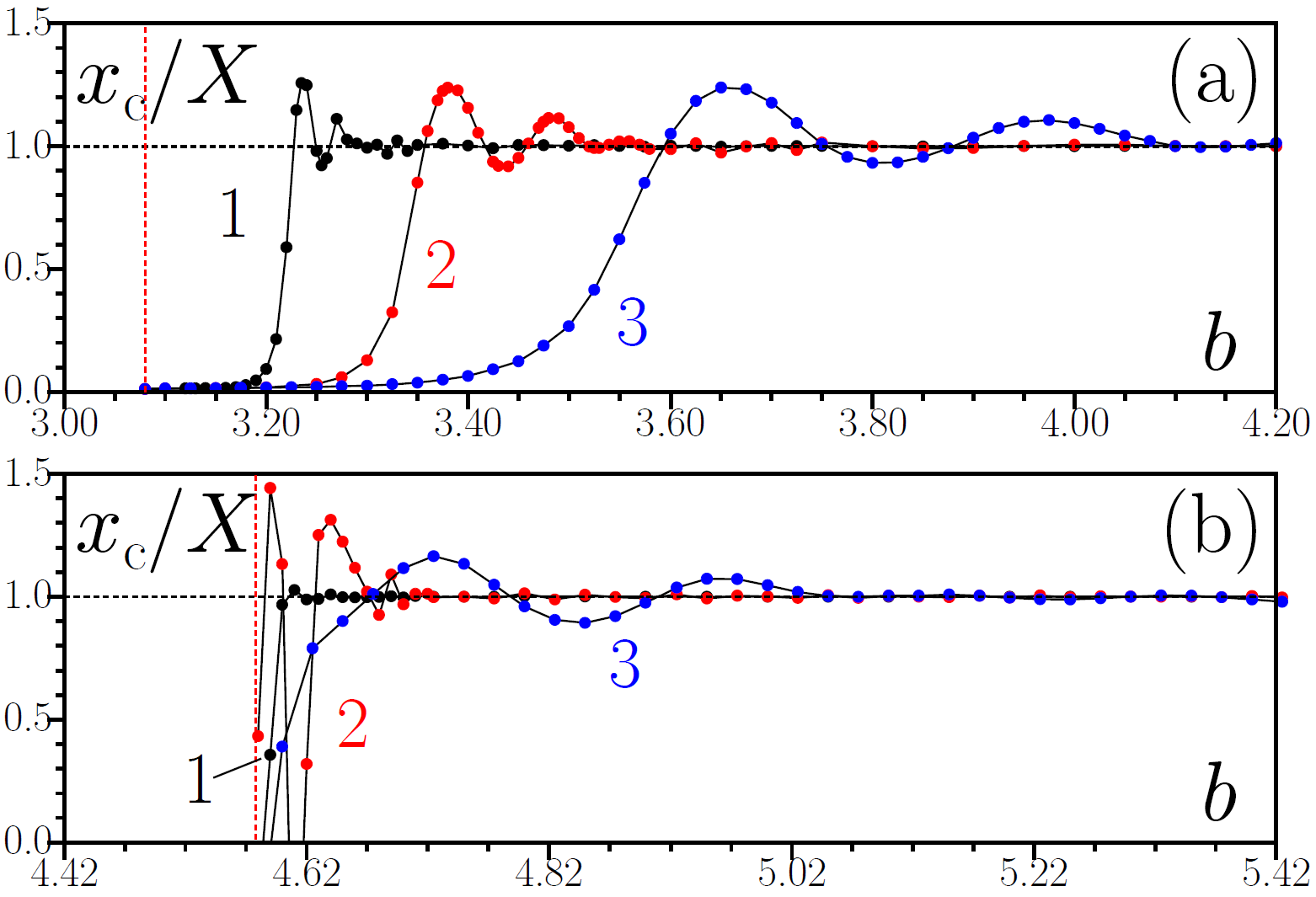}
	 
    \caption{ Soliton center displacement versus propagation constant for different velocities of the sliding lattice $\alpha=0.001$ (curves 1), $\alpha=0.01$ (curves 2) and  $\alpha=0.05$ (curves 3) at $\beta=+3.5$ (a) and $\beta=-2.5$ (b). Vertical red lines correspond to $b_{\rm co}\approx 3.08$ in (a) and $b_{\rm co}\approx 4.58$ in (b). Horizontal dashed lines correspond to $x_c (Z)=X$.}
    \label{fig:four}
\end{figure}

To conclude, we have reported on the pumping of quadratic solitons, which can either be completely suppressed in a non-topological phase, when the input beam has a narrow spectrum, or quantized in a topological phase, when the spectral width of the input beam exceeds the width of the upper bands for each frequency component. The interplay between the band structures of the FF and SH determining the behavior 
of solitons near the propagation constant cutoff, notably affects the transition between non-topological and topological phases that occurs with increase of the soliton propagation constant. It should be mentioned, that parametric interactions between FF and SH waves may lead to the formation of gap solitons with components originating from different gaps, whose pumping was not found in this work, but that could be possible for periodic potentials with different characteristics. Unlike in most previous studies on one-component solitons, particularly cubic solitons, we did not observe breakup or fractional pumping even at very high powers, which remains a unique property of quadratic solitons. \rev{Meantime, the pumping in the presence of  competing quadratic and cubic nonlinearities illustrated in Supplemental Material~\cite{SupplMat}, can display more rich behavior and requires a separate thorough study.}

\acknowledgments
Y.V.K. acknowledges funding by the research project FFUU-2024-0003 of the Institute of Spectroscopy of the Russian Academy of Sciences. V.V.K. was  supported by  the Portuguese Foundation for Science and Technology (FCT) under Contracts UIDB/00618/2020 (DOI: 10.54499/UIDB/00618/2020) and PTDC/FIS-OUT/3882/2020 (DOI: 10.54499/PTDC/FIS-OUT/3882/2020).


\begin{thebibliography}{100}
 
\bibitem{Thouless} D. J. Thouless, Quantization of particle transport. Phys. Rev. B, \textbf{27}, 6083-6087 (1983).

\bibitem{Citro2023} R. Citro and M. Aidelsburger, Thouless pumping and topology, Nat Rev Phys \textbf{5}, 87–101 (2023).


\bibitem{Zilberberg2018} O. Zilberberg, S. Huang, J. Guglielmon, M. Wang, K. P. Chen,Y. E. Kraus, and M. C. Rechtsman, Photonic topological boundary pumping as a probe of 4D quantum Hall physics. Nature \textbf{553}, 59–62 (2018).

\bibitem{Wang2022} P. Wang, Q. Fu, R. Peng, Y. V. Kartashov, L. Torner, V. V. Konotop and F. Ye, Two-dimensional Thouless pumping of light in photonic moiré lattices, Nat. Comm \textbf{13}, 6738 (2022).

\bibitem{Jurgesen2021} M. J\"urgensen, S. Mukherjee, and M. C. Rechtsman, Quantized nonlinear Thouless pumping, Nature {\bf 596}, 63–67 (2021). 


\bibitem{Lohse2016} M. Lohse, C. Schweizer, O. Zilberberg, M. Aidelsburger, I. A. Bloch, Thouless quantum pump with ultracold bosonic atoms in an optical superlattice. Nat. Phys. {\bf 12}, 350–354 (2016). 

\bibitem{Nakajima2016} S. Nakajima, T. Tomita, S. Taie, T. Ichinose, H. Ozawa, L. Wang, M. Troyer and Y. Takahashi, Topological Thouless pumping of ultracold fermions, Nat. Phys. \textbf{12}, 296 (2016).

\bibitem{Taddia2017} L. Taddia, E. Cornfeld, D. Rossini, L. Mazza, E. Sela and R. Fazio, Topological Fractional Pumping with Alkaline-Earth-Like Atoms in Synthetic Lattices, Phys. Rev. Lett. \textbf{118}, 230402 (2017).

 \bibitem{Lohse2018} M. Lohse, C. Schweizer, H. M. Price,O. Zilberberg, I. Bloch. Exploring 4D quantum Hall physics with a 2D topological charge pump. Nature {\bf 553}, 55–58 (2018).

 \bibitem{Nakajima2021} S. Nakajima, N. Takei, K. Sakuma,Y. Kuno, P. Marra and Y. Takahashi, Competition and interplay between topology and quasi-periodic disorder in Thouless pumping of ultracold atoms, Nat. Phys. {\bf 17}, 844–849 (2021).


 \bibitem{Maczewsky2020} L. J. Maczewsky, M. Heinrich, M. Kremer, S. K. Ivanov, M. Ehrhardt, F. Martinez, Y. V. Kartashov, V. V. Konotop, L. Torner, D. Bauer, and A. Szameit, Nonlinearity-induced photonic topological insulator. Science, {\bf 370}, 701 (2020).

\bibitem{Sone2024} K. Sone, M. Ezawa, Y. Ashida, N. Yoshioka, and T. Sagawa, Nonlinearity-induced topological phase transition characterized by the nonlinear Chern number. Nat. Phys. {\bf 20}, 1164 (2024).

\bibitem{KonSal} V. V. Konotop and M. Salerno, Modulational instability in Bose-Einstein condensates in optical lattices, Phys. Rev. A {\bf 65}, 021602(R) (2002).

\bibitem{Bronski2001} J. C. Bronski, L. D. Carr, B. Deconinck, and J. N. Kutz, Bose-Einstein Condensates in Standing Waves: The Cubic Nonlinear Schr\"odinger Equation
with a Periodic Potential. Phys. Rev. Lett. {\bf 86}, 1402 (2001).

\bibitem{Fu2022} Q. Fu, P. Wang, Y. V. Kartashov, V. V. Konotop, and F. Ye, Nonlinear Thouless pumping: solitons and transport breakdown, Phys. Rev. Lett. {\bf 128}, 154101 (2022). 

\bibitem{Fu2022a} Q. Fu, P. Wang, Y. V. Kartashov, V. V. Konotop, and F. Ye, Two-Dimensional Nonlinear Thouless Pumping of Matter Waves, Phys. Rev. Lett. {\bf 129}, 183901 (2022).

\bibitem{Nakagawa2018} M. Nakagawa, T. Yoshida, R. Peters and N. Kawakami, Breakdown of topological Thouless pumping in the strongly interacting regime,
Phys. Rev. B {\bf 98}, 115147 (2018).

\bibitem{Jurgesen2023} M. J\"urgensen, S. Mukherjee, M. C. Rechtsman, C. J\"org, Quantized fractional Thouless pumping of solitons, Nature Phys. {\bf 19},  420–426 (2023).

\bibitem{Kuno2020} Y. Kuno and Y. Hatsugai, Interaction-induced topological charge pump. Phys. Rev. Res. \textbf{2}, 042024(R) (2020).

 

 



\bibitem{Torner2002} L. Torner and A. P. Sukhorukov, Quadratic solitons, Opt. Photonics News {\bf 13}, 42 (2002).

\bibitem{Buryak2002} A. V. Buryak, P. Di Trapani, D. V. Skryabin, and S. Trillo, Optical solitons due to quadratic nonlinearities: From basic physics to futuristic applications, Phys. Rep. {\bf 370}, 63 (2002).


\bibitem{Kartashov2004} Y. V. Kartashov, L. Torner, and V. A. Vysloukh, Multicolor lattice solitons, Opt. Lett. \textbf{29}, 1117 (2004).

\bibitem{Kartashov2021} Y. V. Kartashov, F. Ye, V. V. Konotop, and L. Torner, Multifrequency solitons in commensurate-incommensurate photonic moire lattices, Phys. Rev. Lett. \textbf{127}, 163902 (2021).

\bibitem{Bang1997} O. Bang, P. L. Christiansen, and C. B. Clausen, Stationary solutions and self-trapping in discrete quadratic nonlinear systems, Phys. Rev. E \textbf{56}, 7257 (1997).

\bibitem{Peschel1998} T. Peschel, U. Peschel, and F. Lederer, Discrete bright solitary waves in quadratically nonlinear media, Phys. Rev. E \textbf{57}, 1127 (1998).

\bibitem{Kobyakov1999} A. Kobyakov, S. Darmanyan, T. Pertsch, and F.  Lederer, Stable discrete domain walls and quasi-rectangular solitons in quadratically nonlinear waveguide arrays, J.  Opt. Soc. Am. B \textbf{16}, 1737 (1999).

\bibitem{Sukhorukov2000} A. A. Sukhorukov, Y. S. Kivshar, O. Bang, and C. M. Soukoulis, Parametric localized modes in quadratic nonlinear photonic structures, Phys. Rev. E \textbf{63}, 016615 (2000).

\bibitem{Malomed2002} B. A. Malomed, P. G. Kevrekidis, D. J. Frantzeskakis, H. E. Nistazakis, and A. N. Yannacopoulos, One- and two-dimensional solitons in second-harmonic-generating lattices, Phys. Rev. E \textbf{65}, 056606 (2002).

\bibitem{Iwanow2004} R. Iwanow, R. Schiek, G. I. Stegeman, T. Pertsch, F. Lederer, Y. Min, and W. 
 Sohler, Observation of discrete quadratic solitons, Phys. Rev. Lett. \textbf{93}, 113902 (2004).
 
\bibitem{Xu2005} Z. Xu, Y. V. Kartashov, L.-C. Crasovan, D. Mihalache, and L. Torner, Multicolor vortex solitons in two-dimensional photonic lattices, Phys. Rev. E \textbf{71}, 016616 (2005).

\bibitem{Susanto2007} H. Susanto, P. G. Kevrekidis, R. Carretero-Gonzalez, B. A. Malomed, and D. J. Frantzeskakis, Mobility of discrete solitons in quadratically nonlinear media, Phys. Rev. Lett. \textbf{99}, 214103 (2007).

\bibitem{Setzpfandt2009} F. Setzpfandt, D. N. Neshev, R. Schiek, F. Lederer, A. Tunnermann, and  T.  Pertsch, Competing  nonlinearities in quadratic nonlinear waveguide arrays, Opt. Lett. \textbf{34}, 3589 (2009).

\bibitem{Setzpfandt2010} F. Setzpfandt, A. A. Sukhorukov, D. N. Neshev, R. Schiek, Y. S. Kivshar, and T. Pertsch, Phase transitions of nonlinear waves in quadratic waveguide arrays, Phys.  Rev.  Lett. \textbf{105}, 233905 (2010).

\bibitem{Moreira2012} F. C. Moreira, F. K. Abdullaev, and V. V. Konotop, Gap solitons in nonlinear periodic $\chi^{(2)}$ media. Phys. Rev. A \textbf{85}, 023843 (2012).
 
\bibitem{Moreira2013} F. C. Moreira, V. V. Konotop, and B. A. Malomed, Solitons in $\mathcal{PT}$-symmetric periodic systems with the quadratic nonlinearity. Phys. Rev. A \textbf{87}, 013832 (2013).
 
\bibitem{Kohn1959} W. Kohn, Analytic Properties of Bloch Waves and Wannier Functions, Phys. Rev. \textbf{115}, 809 (1959).

\bibitem{AKKS} G. L. Alfimov, P. G. Kevrekidis, V. V. Konotop, and M. Salerno, Wannier functions analysis of the nonlinear Schrödinger equation with a periodic potential, Phys. Rev. E \textbf{66}, 046608 (2002).

\bibitem{SupplMat} \rev{In the Supplemental Material the cutoff propagation constant, the cascading limit, and the effect of competing nonlinearities are briefly discussed.}

 
\end{thebibliography}
\end{document}